\RequirePackage{lineno}
\documentclass[twocolumn,showpacs,superscriptaddress,amsmath,amssymb]{revtex4-1}
\usepackage{array}
\usepackage{amsfonts}
\usepackage{graphicx}
\usepackage{epsfig}
\usepackage{dcolumn}
\usepackage{bm}
\usepackage{units}
\usepackage{multirow}
\usepackage{bigstrut}
\usepackage{overpic}
\usepackage{color}
\usepackage{lineno}
\usepackage{fmtcount}
\usepackage{relsize}
\usepackage{ulem}
\usepackage{subfigure}

\def \lambdacp   {\Lambda_{c}^{+}}
\def \lambdacm   {\bar{\Lambda}_{c}^{-}}

\def \lamcplamcm {\Lambda_{c}^{+}\bar{\Lambda}_{c}^{-}}

\def \pkpi  {pK^{-}\pi^{+}}
\def \ppipi {p\pi^{+}\pi^{-}}
\def \pkk   {pK^{+}K^{-}}
\def \pphi  {p\phi}
\def \pks   {pK_{S}^{0}}
\def \qqbar {q\bar{q}}

\def \lampi {\Lambda\pi^{+}}
\def \kstar0{K^*(892)^0}

\def \kk    {K^{+}K^{-}}
\def \pipipipi {\pi^{+}\pi^{-}\pi^{+}\pi^{-}}
\def \kkpipi{K^{+}K^{-}\pi^{+}\pi^{-}}
\def \pppipi{p\bar{p}\pi^{+}\pi^{-}}

\def \e    {\varepsilon}
\def \ee   {e^+e^-}
\def \gev  {\mbox{GeV}}

\def \gevcc{\mbox{GeV/$c^2$}}
\def \mev  {\mbox{MeV}}
\def \mevcc{\mbox{MeV/$c^2$}}

\def \ipb  {\mbox{pb$^{-1}$}}
\def \vr   {V_{r}}

\def \BR   {\mathcal{B}}

\def \e    {\varepsilon}

\def \pppm {\pi^{+}\pi^{-}}

\def \pip  {\pi^+}
\def \pim  {\pi^-}

\begin{document}
\title{\boldmath Measurement of Singly Cabibbo Suppressed Decays $\lambdacp\to \ppipi$ and $\lambdacp\to\pkk$}
\author{
   M.~Ablikim$^{1}$, M.~N.~Achasov$^{9,e}$, S. ~Ahmed$^{14}$,
   X.~C.~Ai$^{1}$, O.~Albayrak$^{5}$, M.~Albrecht$^{4}$,
   D.~J.~Ambrose$^{44}$, A.~Amoroso$^{49A,49C}$, F.~F.~An$^{1}$,
   Q.~An$^{46,a}$
% ...
   \\ 
   \vspace{0.2cm}
   (BESIII Collaboration)\\
   \vspace{0.2cm} {\it
     $^{1}$ Institute of High Energy Physics, Beijing 100049, People's Republic of China\\
% ...
$^{4}$ Bochum Ruhr-University, D-44780 Bochum, Germany\\
     $^{5}$ Carnegie Mellon University, Pittsburgh, Pennsylvania 15213, USA\\
% ...
$^{9}$ G.I. Budker Institute of Nuclear Physics SB RAS (BINP), Novosibirsk 630090, Russia\\
% ...
$^{14}$ Helmholtz Institute Mainz, Johann-Joachim-Becher-Weg 45, D-55099 Mainz, Germany\\
% ...
$^{44}$ University of Rochester, Rochester, New York 14627, USA\\
% ...
$^{46}$ University of Science and Technology of China, Hefei 230026, People's Republic of China\\
% ...
$^{49}$ (A)University of Turin, I-10125, Turin, Italy; (B)University of Eastern Piedmont, I-15121, Alessandria, Italy; (C)INFN, I-10125, Turin, Italy\\
% ...
$^{a}$ Also at State Key Laboratory of Particle Detection and Electronics, Beijing 100049, Hefei 230026, People's Republic of China\\
% ...
$^{e}$ Also at the Novosibirsk State University, Novosibirsk, 630090, Russia\\
% ...
    }
  \vspace{0.4cm}
}

\date{\today}
\begin{abstract}
  Using 567 $\ipb$  of data collected with the BESIII detector at a center-of-mass energy of $\sqrt{s}=$ 4.599 $\gev$,  near the $\lamcplamcm$ threshold,  we study the singly Cabibbo-suppressed decays $\lambdacp\to\ppipi$ and $\lambdacp\to\pkk$.
   By normalizing  with respect to the Cabibbo-favored decay $\lambdacp\to\pkpi$, we obtain ratios of branching fractions:  $\frac{\BR(\lambdacp\to\ppipi)} {\BR (\lambdacp\to\pkpi)}$ = $(6.70 \pm 0.48 \pm 0.25)\%$, $\frac{\BR(\lambdacp\to p\phi)}{\BR (\lambdacp\to\pkpi)}$ = $(1.81 \pm 0.33 \pm 0.13)\%$, and $\frac{\BR(\lambdacp\to\pkk_{\text{non-}\phi})}{\BR (\lambdacp\to\pkpi)}$ = $(9.36 \pm 2.22 \pm 0.71)\times10^{-3}$,
  where the uncertainties are statistical and systematic, respectively.  The absolute branching fractions are also presented.  Among these measurements,  the decay $\lambdacp\to\ppipi$ is observed for the first time, and the precision of the branching fraction for $\lambdacp\to\pkk_{\text{non-}\phi}$ and $\lambdacp\to p\phi$ is significantly improved.
\end{abstract}


\maketitle

Hadronic decays of charmed baryons provide an ideal laboratory to understand
the interplay of weak and strong interaction in the charm region~\cite{
2015Yhaiyang, 2016Ylucd, 1998YKohara, 1998Mikhail, 1997KKSharma, 1994TUppal,
1994PZenczykowsky, 1992JGKorner, 1979JGKorner}, which is complementary to charmed mesons.
They also provide essential input for studying the decays of $b$-flavored
hadrons involving a $\Lambda_c$ in the final state~\cite{2015WDetmold, 2016RDutta}.
In contrast to the charmed meson decays, which are usually dominated
by factorizable amplitudes,
decays of charmed baryons receive sizable nonfactorizable contributions from $W$-exchange diagrams, which
are subject to color and helicity suppression.
The study of nonfactorizable contributions is critical to understand the dynamics of charmed baryon decays.

Since the first discovery of the ground state charmed baryon $\Lambda_c$ in 1979~\cite{1980GSAbrams,1979AMCnops},
progress with charmed baryons has been relatively slow, due to a scarcity of experimental data.
Recently,  based on an $\ee$ annihilation data sample of 567 $\ipb$~\cite{2015MAblikimLumi}
at a center-of-mass (c.m.) energy of $\sqrt{s}= 4.599\;\gev$,
the BESIII Collaboration measured the absolute branching fractions (BFs) of 12
Cabibbo-favored (CF) $\lambdacp$  hadronic decays  with a significantly improved
precision~\cite{lambdcpBr}.
For many other CF charmed baryon decay modes and most of the singly Cabibbo-suppressed (SCS) decays, however, no precision
measurements are available;
many of them even have not yet been measured~\cite{2014PDG}.
As a consequence, we are not able to distinguish between the theoretical predictions among
the different models~\cite{1998YKohara, 1998Mikhail, 1997KKSharma, 1994TUppal, 1994PZenczykowsky, 1992JGKorner, 1979JGKorner}.

The SCS decays $\lambdacp\to\ppipi$ and $\lambdacp\to\pkk$ proceed via the external $W$-emission, internal $W$-emission and $W$-exchange processes. 
Precisely measuring and comparing their BFs may help to reveal the $\Lambda_c$ internal dynamics~\cite{2015Yhaiyang}.
A measurement of the SCS mode $\lambdacp\to p \phi$ is of particular interest because it receives contributions only from
the internal $W$-emission diagrams, which can reliably be obtained by a factorization approach~\cite{2015Yhaiyang}.
An improved measurement of the $\lambdacp\to\pphi$ BF is thus essential to validate theoretical models and test the application of large-$N_{c}$ factorization in the charmed baryon sector~\cite{2010Yhaiyang},
where, $N_{c}$ is the number of colors.

In this Letter, we describe a search for the SCS decays $\lambdacp\to\ppipi$ and present an improved measurement of
the $\lambdacp\to\pkk_{\text{non-}\phi}$ and $\lambdacp\to p \phi$ BFs.
The BFs are measured relative to the CF mode $\lambdacp\to\pkpi$.
Our analysis is based on the same data sample as that used in Ref.~\cite{lambdcpBr} collected by the BESIII detector.
Details on the features and capabilities of the BESIII detector can be found in Ref.~\cite{2009MAblikimDet}.
Throughout this Letter, charge-conjugate modes are implicitly included, unless otherwise stated.

The GEANT4-based~\cite{GEANT} Monte Carlo (MC) simulations of $\ee$ annihilations
are used to understand the backgrounds and to estimate detection efficiencies.
The generator $\textsc{KKMC}$~\cite{2001SJadach} is used to simulate the
beam-energy spread and initial-state radiation (ISR) of the $\ee$ collisions.
The inclusive MC sample includes $\lambdacp\lambdacm$ events,
charmed meson $D^{(*)}_{(s)}$ pair production, ISR returns to lower-mass $\psi$
states, and continuum processes $\ee\to\qqbar$  ($q = u, d, s$).
Decay modes as specified in the PDG~\cite{2014PDG} are modeled with
EVTGEN~\cite{2008RGPing,2001DJLange}.
Signal MC samples of $\ee\to\lambdacp\lambdacm$ are produced in which the
$\lambdacp$ decays to the interested final state ($\pkpi$, $\ppipi$, or $\pkk$)
together with the $\lambdacm$ decaying generically to all possible final states.

Charged tracks are reconstructed from hits in the main drift chamber (MDC) and are required to have polar angles within $|\cos\theta|<0.93$.
The points of closest approach of the charged tracks to the interaction point (IP) are required to be within 1 cm in the plane perpendicular to the beam ($V_r$) and $\pm$10 cm along the beam ($V_z$).
Information from the time-of-flight (TOF) system and $dE/dx$ in the MDC are combined to form PID confidence levels (C.L.) for the $\pi$, $K$ and $p$ hypotheses.
Each track is assigned to the particle type with the highest particle identification (PID) C.L.
To avoid backgrounds from beam interactions with residual gas or detector materials (beam pipe and MDC inner wall), a further requirement of $\vr<0.2$~cm is imposed for the proton.

$\lambdacp$ candidates are reconstructed by considering all combinations of charged tracks in the final states of interest $\pkpi$, $\ppipi$, and $\pkk$.
Two variables, the energy difference $\Delta E = E - E_{\text{beam}}$ and the beam-constrained mass $M_{\text{BC}} = \sqrt{E_{\text{beam}}^2/c^4 - p^2/ c^2}$, are used to identify the $\lambdacp$ candidates.
Here, $E_{\text{beam}}$ is the beam energy, and $E(p)$ is the reconstructed energy (momentum) of the $\Lambda^+_c$ candidate in the $e^+e^-$ c.m. system.
A $\Lambda^+_c$ candidate is accepted with  $M_{\rm BC}>2.25~\gevcc$ and $|\Delta E|<20$ MeV (corresponding to 3 times the resolution).
For a given signal mode, we accept only one candidate per $\Lambda_c$ charge per event.
If multiple candidates are found, the one with the smallest $|\Delta E|$ is selected.
The $\Delta E$ sideband region, $40 <|\Delta E|< 60 \;\mev$, is defined to investigate potential backgrounds.

For the $\lambdacp\to\ppipi$ decay, we reject $K^0_S$ and $\Lambda$ candidates by requiring $|M_{\pip\pim}-M^{\rm PDG}_{K^0_{S}}|>15$~$\mevcc$ and $|M_{p\pim}-M^{\rm PDG}_{\Lambda}|>$ 6 $\mevcc$, corresponding to 3 times
the resolution, where $M^{\rm PDG}_{K^0_{S}}$ ($M^{\rm PDG}_{\Lambda}$) is the $K^0_S$ ($\Lambda$) mass quoted from the PDG~\cite{2014PDG} and $M_{\pip\pim}$ ($M_{p\pim}$) is the $\pip\pim$ ($p\pim$) invariant mass.
These requirements suppress the peaking backgrounds of the CF decays $\lambdacp\to \Lambda \pip$ and $\lambdacp\to p K^0_S$,
which have the same final state as the signal.

\begin{figure}[htbp]
  \centering
  \mbox{
    \begin{overpic}[width=0.22\textwidth, height=0.18\textwidth]{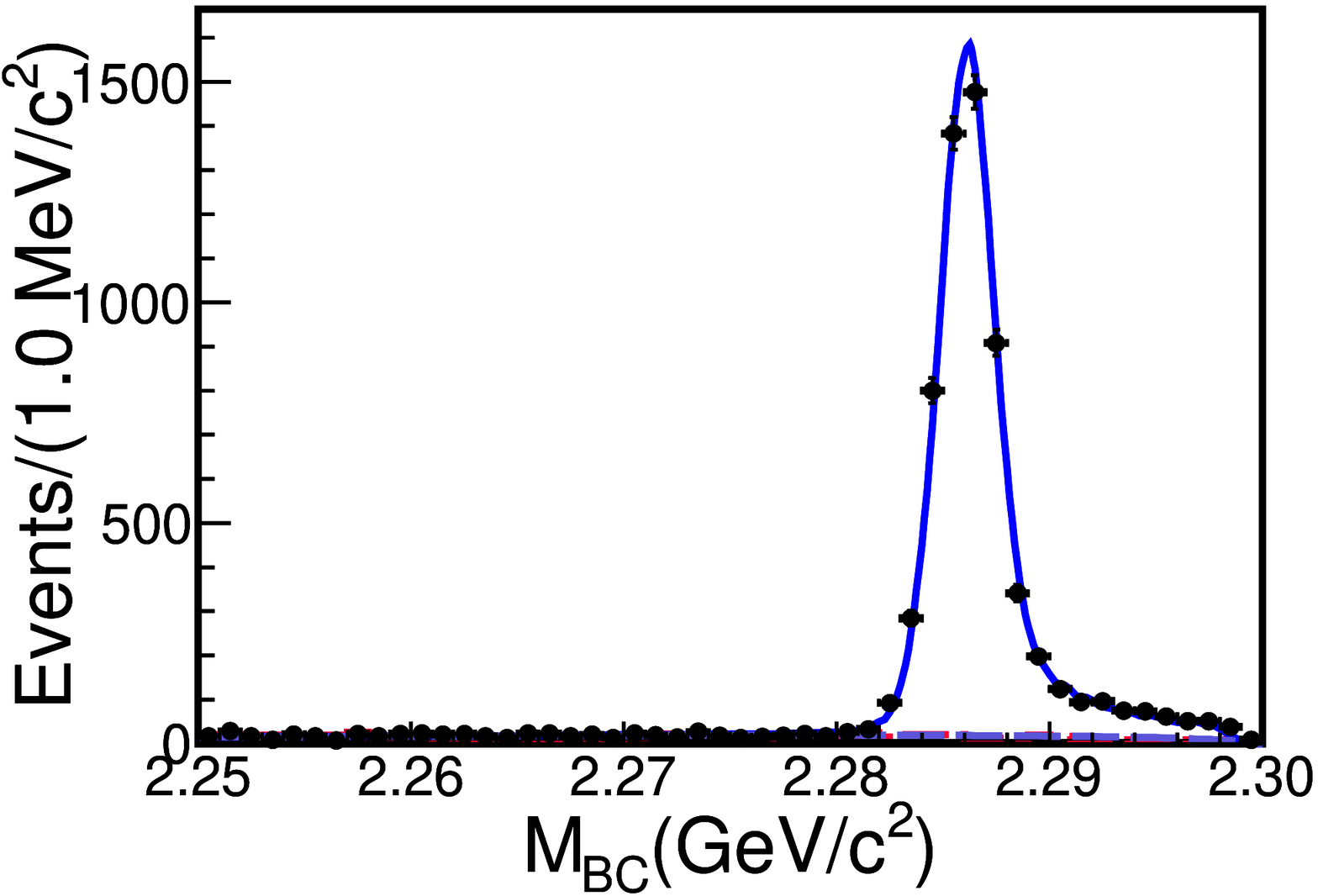}
    \put(80,70){$(a)$}
    \end{overpic}
    \begin{overpic}[width=0.22\textwidth, height=0.18\textwidth]{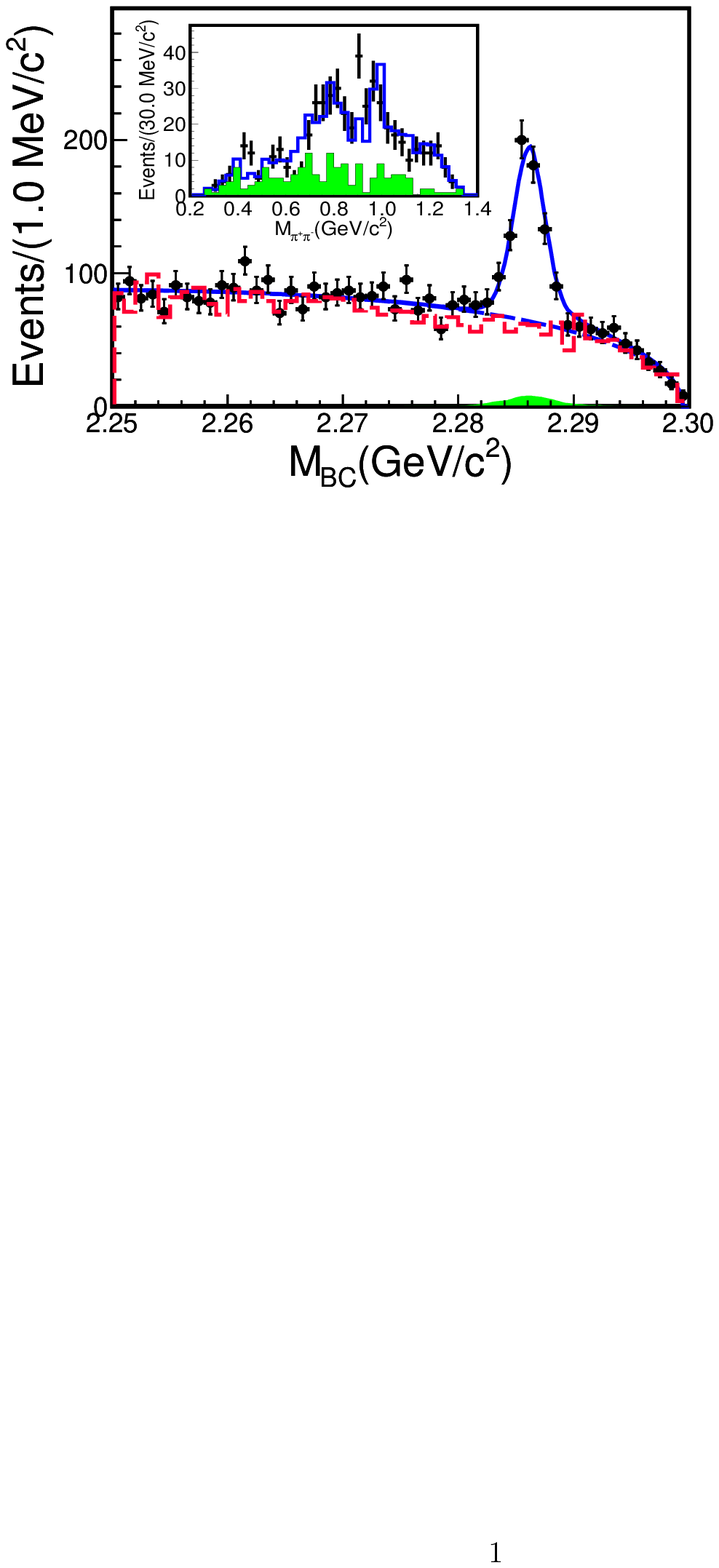}
    \put(80,70){$(b)$}
    \end{overpic}
  }
  \caption{Distributions of $M_{\rm BC}$ for the decays (a) $\lambdacp\to\pkpi$
	and (b) $\lambdacp\to\ppipi$.
	Points with an error bar are data, the blue solid lines show the total fits,
	the blue long dashed lines are the combinatorial background shapes,
    and the red long dashed histograms are data from the $\Delta E$ sideband
	region for comparison.
	In (b), the green shaded histogram is the peaking background from the CF decays $\lambdacp\to\pks$ and
    $\lambdacp\to\lampi$.
    The inset plot in (b) shows the $\pppm$ invariant mass distribution with the additional requirement $|\Delta E|<8$ MeV and $2.2836 < M_{\rm BC}< 2.2894~\gevcc$, where the dots with an error bar are for the data, the blue solid histogram shows the fit curve from PWA, and the green shaded histogram shows background estimated from the $M_{\rm BC}$ sideband region.
 }
  \label{fig:fitmbc}
\end{figure}

\begin{figure}[htbp]
  \mbox{
    \begin{overpic}[width=0.22\textwidth, height=0.18\textwidth]{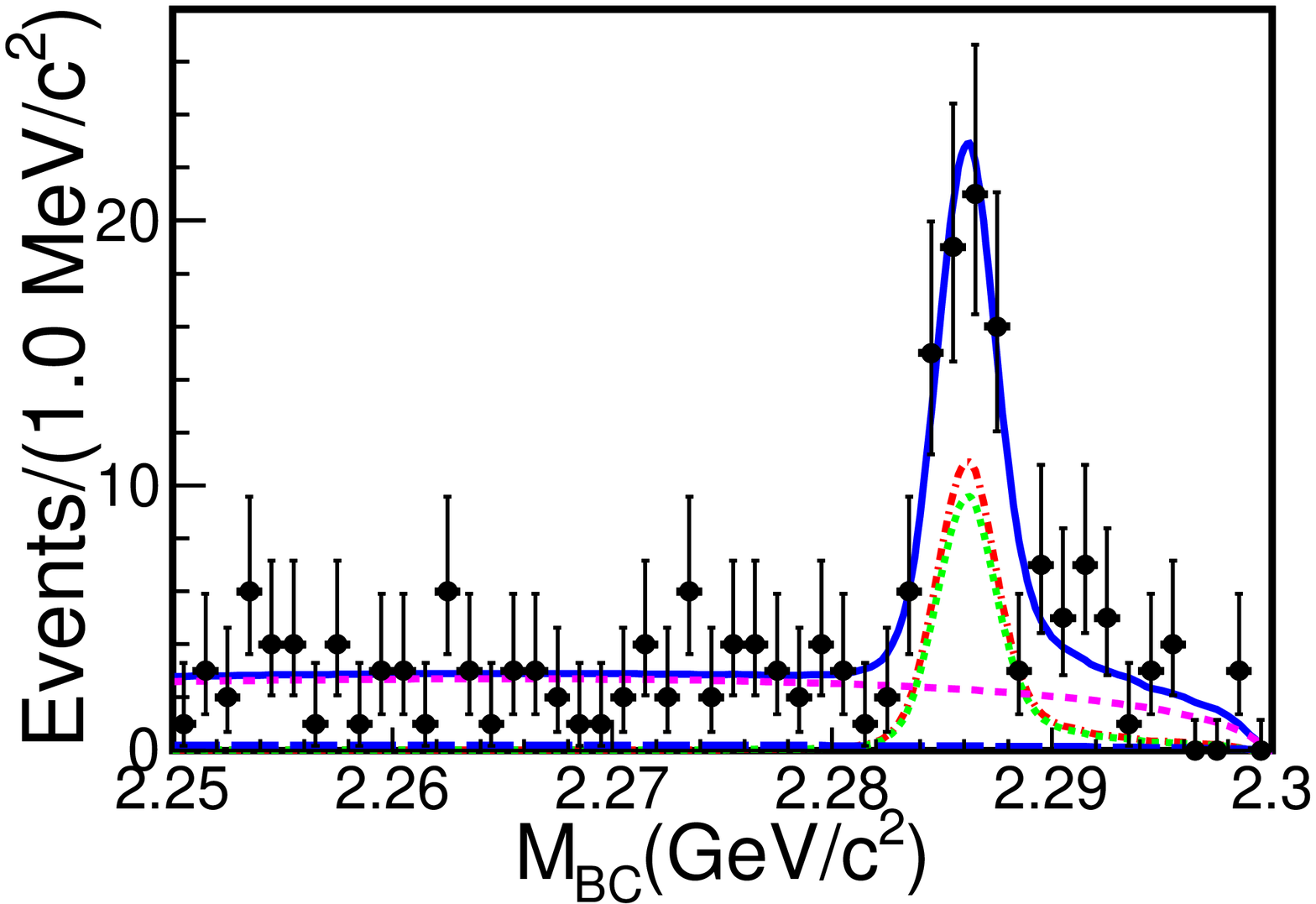}
    \put(80,70){$(a)$}
    \end{overpic}
    \begin{overpic}[width=0.22\textwidth, height=0.18\textwidth]{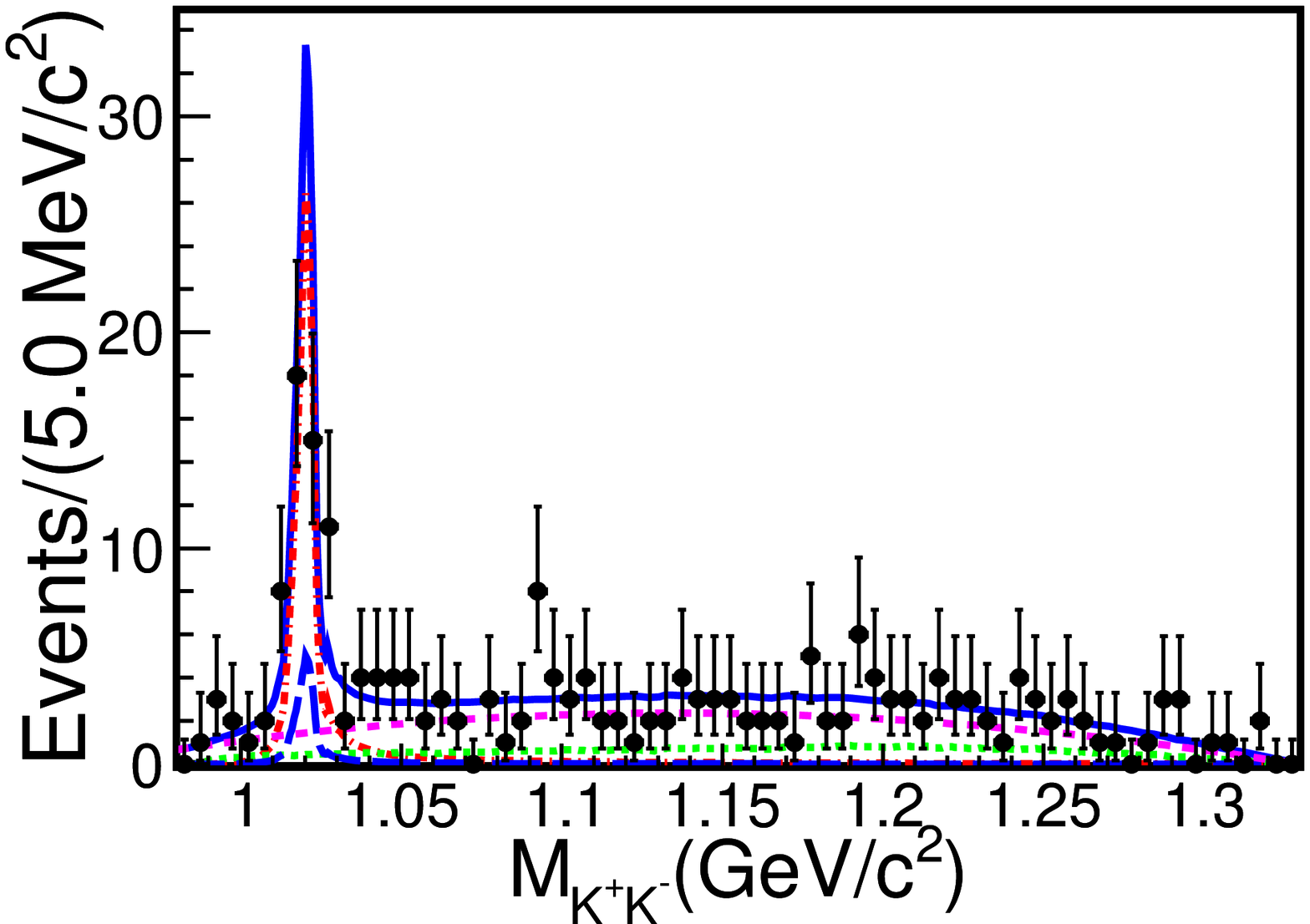}
    \put(80,70){$(b)$}
    \end{overpic}
  }
  \mbox{
    \begin{overpic}[width=0.22\textwidth, height=0.18\textwidth]{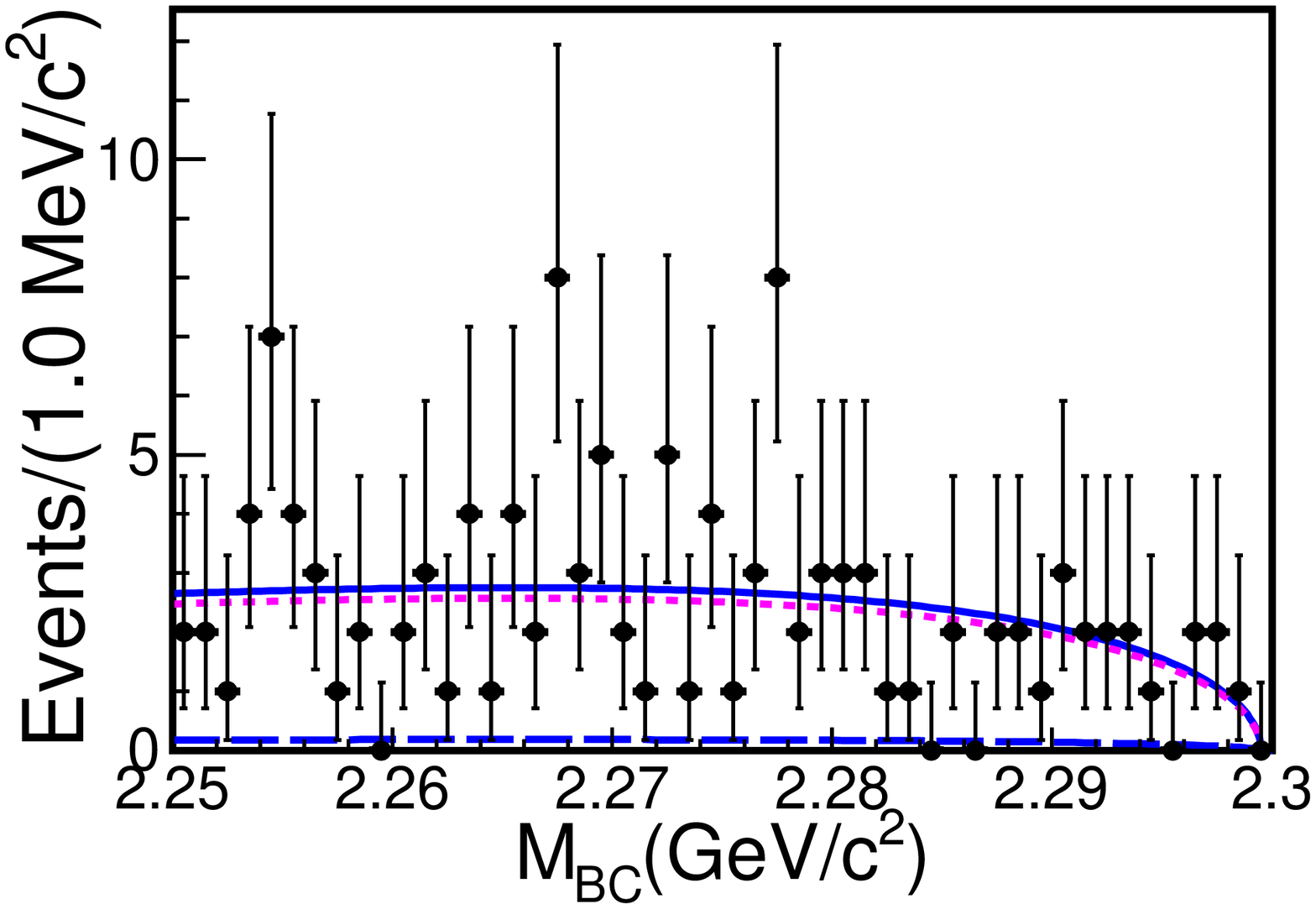}
    \put(80,70){$(c)$}
    \end{overpic}
    \begin{overpic}[width=0.22\textwidth, height=0.18\textwidth]{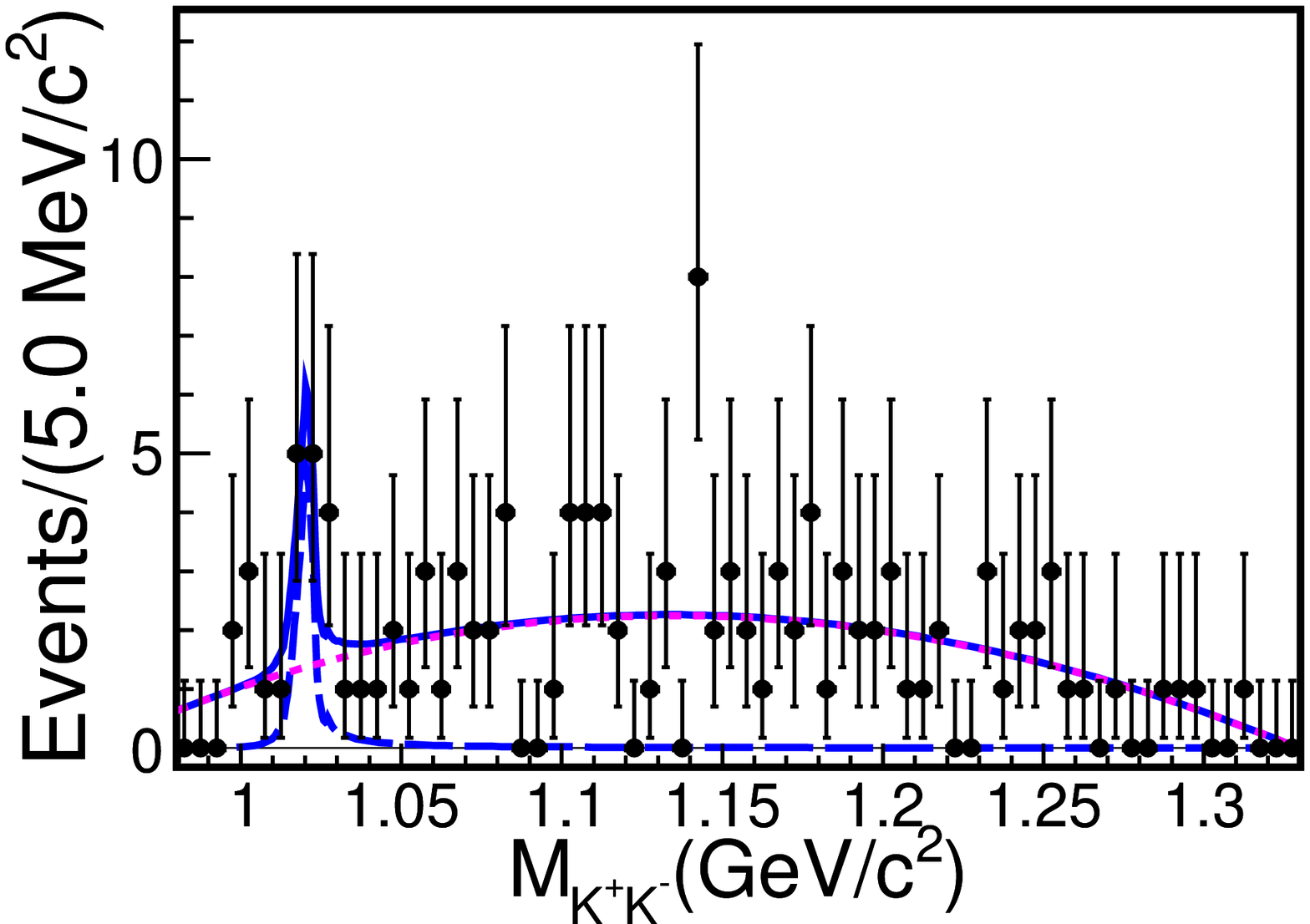}
    \put(80,70){$(d)$}
    \end{overpic}
  }
  \caption{Distributions of $M_{\rm BC}$ (left) and $M_{\kk}$ (right) for
	data in the $\Delta E$ signal region (upper) and sideband region (bottom) for the decay
    $\lambdacp\to \pkk$.
    The blue solid curves are for the total fit results, the red dash-dotted curves show the
    $\lambdacp\to\pphi\to p\kk$ signal, the green dotted curves show the
	$\lambdacp\to\pkk_{\text{non-}\phi}$ signal, the blue long-dashed curves are the
	background with $\phi$ production, and the magenta dashed curves are the non-$\phi$
	background.
          }
  \label{fig:phi-mbc-2Dfit}
\end{figure}

With the above selection criteria, the $M_{\rm BC}$ distributions are depicted in Fig.~\ref{fig:fitmbc} for the decays $\lambdacp\to\pkpi$ and $\lambdacp\to\ppipi$  and in Fig.~\ref{fig:phi-mbc-2Dfit} (a) for the decay $\lambdacp\to\pkk$.
Prominent $\lambdacp$ signals are observed.
The inclusive MC samples are used to study potential backgrounds.
For the decays $\lambdacp\to\pkpi$ and $\lambdacp\to\pkk$, no peaking background is evidenced in the $M_{\rm BC}$ distributions, while for the decay $\lambdacp\to\ppipi$, the peaking backgrounds of $28.2\pm1.6$ events from the decays $\lambdacp\to \Lambda \pip$ and $\lambdacp\to p K^0_S$ are expected, where the uncertainty comes from the measured BFs in Ref.~\cite{lambdcpBr}.
The cross feed between the decay modes is negligible by the MC studies.

To obtain the signal yields of the decays $\lambdacp\to\pkpi$ and $\lambdacp\to\ppipi$, a maximum likelihood fit is performed to the corresponding $M_{\rm BC}$ distributions.
The signal shape is modeled with the MC simulated shape convoluted with a Gaussian function representing the resolution difference and potential mass shift between the data and MC simulation.
The combinatorial background is modeled by an ARGUS function~\cite{1990argus}.
In the decay $\lambdacp\to \ppipi$, the peaking background is included in the fit, and is modeled with the MC simulated shape convoluted with the same Gaussian function for the signal, while the magnitude is fixed to the MC prediction.
The fit curves are shown in Fig.~\ref{fig:fitmbc}.
The $M_{\rm BC}$ distribution for events in the $\Delta E$ sideband region is also shown in Fig.~\ref{fig:fitmbc}(b), and a good agreement with the fitted background shape is indicated.
The signal yields are summarized in Table~\ref{tab:SigEff}.

For the decay $\lambdacp\to\pkk$, a prominent $\phi$ signal is observed in the $M_{\kk}$ distribution, as shown in Fig.~\ref{fig:phi-mbc-2Dfit} (b).
To determine the signal yields via $\phi$ ($N^{\phi}_{\rm sig}$) and non-$\phi$ ($N^{\text{non-}\phi}_{\text{sig}}$) processes and to better model the background,  we perform a two-dimensional unbinned extended maximum likelihood fit to the $M_{\rm BC}$ versus $M_{\kk}$ distributions for events in the $\Delta E$ signal region and sideband region simultaneously.
In the $M_{\rm BC}$ distribution, the shapes of $\Lambda_c$ signal (via $\phi$ or non-$\phi$ process) and background, denoted as $S_{M_{\rm BC}}$ and $B_{M_{\rm BC}}$, respectively, are modeled similarly to those in the decay $\lambdacp\to\ppipi$.
In the $M_{\kk}$ distribution, the $\phi$ shape for the $\Lambda_c$ process ($\lambdacp\to p\phi\to\pkk$), $S^{\phi}_{M_{\rm KK}}$, is modeled with a relativistic Breit-Wigner function convoluted with a Gaussian function representing the detector resolution, while that for the $\Lambda_c$ decay without $\phi$ ($\lambdacp\to\pkk$), $S^{\text{non-}\phi}_{M_{\rm KK}}$, is represented by the MC shape with a uniform distribution in $\kk$ phase space.
The shape for the non-$\Lambda_c$ background including $\phi$ state, $B^{\phi}_{M_{\rm KK}}$, has the same parameters as $S^{\phi}_{M_{\rm KK}}$, while that for the background without $\phi$, $B^{\text{non-}\phi}_{M_{\rm KK}}$, is described by a third-order polynomial function.
Detailed MC studies indicate the non-$\Lambda_c$ background (both with and without $\phi$ included) have the same shapes and yields in both the $\Delta E$ signal and sideband regions, where the yields are denoted as $N^{\phi}_{\rm bkg}$ and $N^{non-\phi}_{\rm bkg}$, respectively.
The likelihoods for the events in the $\Delta E$ signal and sideband regions are given in Eqs.~\eqref{eq:pdf:2d:sig}~and~\eqref{eq:pdf:2d:side}, respectively:

 \begin{eqnarray}
\mathcal{L_{\rm sig}} &=& \frac{e^{-(N^{\phi}_{\rm sig}+N^{\text{non-}\phi}_{\rm sig} +N^{\phi}_{\rm bkg}+N^{\text{non-}\phi}_{\rm bkg} )}}{N_\text{sig}!}\nonumber\\
&&\times \prod^{N_\text{sig}}_{i=1}[ N^{\phi}_{\rm sig} S_{M_{\rm BC}}(M^i_{\rm BC})\times
{S}^{\phi}_{M_{\rm KK}}(M^i_{\kk}) \nonumber\\
&&+  N^{\text{non-}\phi}_{\rm sig} {S}_{M_{\rm BC}} (M^i_{\rm BC})\times
{S}^{\text{non-}\phi}_{M_{\rm KK}}(M^i_{\kk}) \nonumber\\
&&+ N^{\phi}_{\rm bkg} {B}_{M_{\rm BC}} (M^i_{\rm BC})\times {B}^{\phi}_{M_{\rm KK}}(M^i_{\kk})\nonumber \\
&&+ N^{\text{non-}\phi}_{\rm bkg} {B}_{M_{\rm BC}} (M^i_{\rm BC})\times {B}^{\text{non-}\phi}_{M_{\rm KK}} (M^i_{\kk})]
,  \label{eq:pdf:2d:sig}
\end{eqnarray}

\begin{eqnarray}
\mathcal{L_{\rm side}} &=& \frac{e^{-(N^{\phi}_{\rm bkg} + N^{\text{non-}\phi}_{\rm bkg})}}{N_\text{side}!} \nonumber\\
&&\times \prod^{N_\text{side}}_{i=1}[N^{\phi}_{\rm bkg} {B}_{M_{\rm BC}} (M^i_{\rm BC})\times {B}^{\phi}_{M_{\rm KK}}(M^i_{\kk})\nonumber \\
&&+ N^{\text{non-}\phi}_{\rm bkg} {B}_{M_{\rm BC}} (M^i_{\rm BC})\times {B}^{\text{non-}\phi}_{M_{\rm KK}} (M^i_{\kk})]
,  \label{eq:pdf:2d:side}
\end{eqnarray}

where the parameter $N_\text{sig}$ ($N_\text{side}$) is the total number of selected candidates in the $\Delta E$ signal (sideband) region and $M^i_{\rm BC}$ and $M^i_{\kk}$ are the values of $M_{\rm BC}$ and $M_{\kk}$, respectively, for the $i$th event.
We use the product of PDFs, since the $M_{\rm BC}$ and $M_{\kk}$ are verified to be uncorrelated for each component by MC simulations.

The signal yields are extracted by minimizing the negative log-likelihood $-\ln\mathcal{L}$ = $(-\ln\mathcal{L_{\rm sig}})+(-\ln\mathcal{L_{\rm side}})$.
The fit curves are shown in Fig.~\ref{fig:phi-mbc-2Dfit}, and the yields are listed in Table~\ref{tab:SigEff}.
The significance is estimated by comparing the likelihood values with and without the signal components included, incorporating with the change of the number of free parameters,  listed in Table~\ref{tab:SigEff}.

\begin{table}[htbp]
  \begin{center}
  \caption{Summary of signal yields in data ($N_{\rm signal}$), detection efficiencies ($\e$), and the significances.
           The errors are statistical only.}
  \begin{tabular}{lcccc}
      \hline \hline
      Decay modes                     &  $N_{\rm signal}$  &  $\e(\%)$      &  Significance   \\ \hline
      $\lambdacp\to\pkpi$             &  $5940\pm85$       &  $48.0\pm0.1$  &  -              \\
      $\lambdacp\to\ppipi$            &  $495 \pm35$       &  $59.7\pm0.1$  &  $16.2 \sigma$  \\
      $\lambdacp\to\pkk$(via~$\phi$)  &  $44 \pm 8$        &  $40.2\pm0.1$  &  $9.6 \sigma$   \\
      $\lambdacp\to\pkk$(non-$\phi$)  &  $38 \pm 9$        &  $32.7\pm0.1$  &  $5.4 \sigma$   \\
      \hline\hline
  \end{tabular}
  \label{tab:SigEff}
  \end{center}
\end{table}

In the decays $\lambdacp \to \pkpi$ and $\lambdacp\to\ppipi$, the detection efficiencies are estimated with data-driven MC samples generated according to the results of a simple partial wave analysis (PWA) by the covariant helicity coupling amplitude ~\cite{amplitude1,amplitude2} for the quasi-two-body decays.
In the decay $\lambdacp\to\ppipi$, prominent structures arising from $\rho^0(770)$ and $f_0(980)$ resonances are observed in the $M_{\pppm}$ distribution as shown in the inset plot of Fig.~\ref{fig:fitmbc}(b) and are included in the PWA.
Because of the limited statistics and relatively high background, the PWA does not allow for a reliable extraction of BFs for intermediate states; it however does describe the kinematics well, and it is reasonable for the estimation of the detection efficiency. The corresponding uncertainty is taken into account as a systematic error.
For the decays $\lambdacp\to\pkk$ via $\phi$ or non-$\phi$, the detection efficiencies are estimated with phase space MC samples, where the angular distribution of the decay $\phi\to\kk$ is considered.

We measure the relative BFs of the SCS decays with respect to that of the CF decay $\lambdacp\to\pkpi$ and the absolute BFs by incorporating $\BR(\lambdacp\to\pkpi)=(5.84 \pm 0.27 \pm 0.23)\%$ from the most recent BESIII measurement~\cite{lambdcpBr}.
Several sources of systematic uncertainty, including tracking and PID efficiencies and the total number of $\lambdacp\lambdacm$ pairs in the data, cancel when calculating the ratio of BFs,
due to the similar kinematics between the SCS and CF decays.
When calculating these uncertainties, cancellation has been taken into account whenever possible.

\begin{table}[htbp]
  \begin{center}
  \footnotesize
  \caption{The systematic uncertainties (in percent) in the relative BF measurements.
           The uncertainty of the reference BF ${\cal B}_{\rm ref}$ applies only to the absolute BF measurements.}
  \begin{tabular}{c c c c}
      \hline \hline
      Sources                   & $\lambdacp\to\ppipi$ & $\lambdacp\to p \phi$ & $\lambdacp\to\pkk_{\text{non-}\phi}$  \\ \hline
      Tracking                  & $1.1$         & $2.6$                  & $1.6$                 \\
      PID                       & $1.3$         & $1.5$                  & $1.9$                 \\
      $\vr$ requirement         & $0.6$         & $2.5$                  & $2.5$                 \\
      $K^0_S$/$\Lambda$ vetoes  & $0.7$         & $-$                    & $-$                   \\
      $\Delta E$ requirement    & $0.5$         & $0.7$                  & $0.9$                 \\
      Fit                       & $2.7$         & $5.8$                  & $6.6$                 \\
      Cited branching ratio     & $-$           & $1.0$                  & $-$                   \\
      MC model                  & $1.4$         & $1.0$                  & $1.1$    \\
      MC statistics             & $0.3$         & $0.4$                  & $0.4$                 \\ \hline
      Total                     & $3.7$         & $7.2$                  & $7.6$                 \\ \hline
      ${\cal B}_{\rm ref}$      & $6.1$         & $6.1$                  & $6.1$                 \\
      \hline\hline
  \end{tabular}
  \label{tab:uncertainties}
  \end{center}
\end{table}

The uncertainties associated with tracking and PID efficiencies for $\pi$, $K$, and proton are studied as a function of (transverse) momentum with samples
of $\ee\to\pipipipi$, $\kkpipi$ and $\pppipi$ from data taken at $\sqrt{s} > 4.0\;\gev$.
To extract the tracking efficiency for particle $i$ ($i$ = $\pi$, $K$, or ptoton), we select the corresponding samples by missing particle $i$ with high purity, and the ratio to find the track $i$ around the missing direction is the tracking efficiency.
Similarly, we select the control sample without a PID requirement for particle $i$, and then the PID requirement is further implemented.
The PID efficiency is the ratio between the number of candidates with and without the PID requirement.
The differences on the efficiency between the data and MC simulation weighted by the (transverse) momentum according to the data are assigned as uncertainties.

The uncertainties due to the $\vr$ requirements and $K^0_S$/$\Lambda$ vetoes (in $\lambdacp\to\ppipi$ only) are investigated by repeating the analysis with alternative requirements ($\vr <$ 0.25 cm, $|M_{\pip\pim}-M^{\rm PDG}_{K^0_{S}}|>20$~$\mevcc$, and
$|M_{p\pim}-M^{\rm PDG}_{\Lambda}|>$ 8 $\mevcc$, respectively).
The resulting differences in the BFs are taken as the uncertainties.
Uncertainties related to the $\Delta E$ resolution are estimated by widening the $\Delta E$ windows from $3\sigma$ to $4\sigma$ of the resolution.

For the decays $\lambdacp\to\pkpi$ and $\lambdacp\to\ppipi$, the signal yields are determined from fits to the $M_{\rm BC}$ distributions.
Alternative fits are carried out by varying the fit range, signal shape, background shape and the expected number of peaking backgrounds.
The resultant changes in the BFs are taken as uncertainties.
In the decay $\lambdacp\to\pkk$,
the uncertainties associated with the fit are studied by varying the fit ranges, signal and background shapes
for both the $M_{\rm BC}$ and $M_{\kk}$ distributions, and $\Delta E$ sideband region.

The following four aspects are considered for the MC simulation model uncertainty.
(a) The uncertainties related to the beam energy spread are investigated by changing its value in the simulation by $\pm0.4$ MeV, where the nominal values is 1.5 $\mev$ determined by the data.
The larger change in the measurement is taken as a systematic uncertainty.
(b) The uncertainties associated with the input line shape of the $\ee\to\lambdacp\lambdacm$ cross section is estimated by replacing the line shape directly from BESIII data with that from Ref.~\cite{belleLambda}.
(c) The $\lambdacp$ polar angle distribution in the $\ee$ rest frame is parameterized with $1+\alpha\cos^2\theta$, where the $\alpha$ value is extracted from the data.
The uncertainties due to the $\lambdacp$ polar angle distribution are estimated by changing the $\alpha$ value by one standard deviation.
(d) The decays $\lambdacp\to\pkpi$ and $\lambdacp\to\ppipi$ are modeled by a data-driven method according to PWA results.
The corresponding uncertainties are estimated by changing the intermediate states included, changing the parameters of
the intermediate states by one standard deviation quoted in the PDG~\cite{2014PDG}, and varying the background treatment in the PWA and the output parameters for the coupling.
Assuming all of the above PWA uncertainties are independent, the uncertainty related to MC modeling is the quadratic sum of all individual values.
For the non-$\phi$ decay $\lambdacp\to\pkk$, phase space MC samples with an $S$ wave for the $\kk$ pair are used to estimate the detection efficiency.
An alternative MC sample with a $P$ wave between the $\kk$ pair is also used,
and the resultant difference in efficiency is taken as the uncertainty.
The uncertainties due to limited MC statistics in both the measured and reference modes are taken into account.
	
Assuming all uncertainties,  summarized in Table~\ref{tab:uncertainties}, are independent, the total uncertainties in the relative BF measurements are obtained by adding the individual uncertainties in quadrature.
For the absolute BF measurements,  the uncertainty due to the reference BF ${\cal B}_{\rm ref} (\lambdacp\to\pkpi)$, listed in Table~\ref{tab:uncertainties}, too, is included.

\begin{table*}[htbp]
  \begin{center}

  \caption{Summary of relative and absolute BFs, and comparing with the results from PDG~\cite{2014PDG}.
           Uncertainties are statistical, experimental systematic, and reference mode uncertainty, respectively.
          }
  \begin{tabular}{l c c}
      \hline \hline
    Decay modes                           &  ${\cal B}_{\rm mode}/{\cal B}_{\rm ref}$ (This work)  &  ${\cal B}_{\rm mode}/{\cal B}_{\rm ref}$ (PDG average)   \\ \hline
    $\lambdacp\to\ppipi$                  &  $(6.70 \pm 0.48 \pm 0.25)\times 10^{-2}$              &  $(6.9\pm3.6)\times 10^{-2}$                              \\
    $\lambdacp\to p \phi $                &  $(1.81 \pm 0.33 \pm 0.13)\times 10^{-2}$              &  $(1.64\pm0.32)\times 10^{-2}$                            \\
    $\lambdacp\to\pkk$ ($\text{non-}\phi$)&  $(9.36 \pm 2.22 \pm 0.71)\times 10^{-3}$              &  $(7\pm2\pm2)\times 10^{-3}$                              \\ \hline \hline
    $~~~~~~-$                             &  ${\cal B}_{\rm mode}$ (This work)                     &  ${\cal B}_{\rm mode}$ (PDG average)                      \\ \hline
    $\lambdacp\to\ppipi$                  &$(3.91\pm0.28\pm0.15\pm0.24)\times10^{-3}$              &~~$(3.5 \pm 2.0)\times 10^{-3}$~~                          \\
    $\lambdacp\to p \phi $                & $(1.06\pm0.19\pm0.08\pm0.06)\times10^{-3}$             &  $(8.2 \pm 2.7)\times 10^{-4}$                            \\
    $\lambdacp\to\pkk$ ($\text{non-}\phi$)&$(5.47\pm1.30\pm0.41\pm0.33)\times10^{-4}$              &  $(3.5 \pm 1.7)\times 10^{-4}$                            \\
    \hline\hline
  \end{tabular}
  \label{tab:br}
  \end{center}
\end{table*}

In summary, based on 567 $\ipb$ of $\ee$ annihilation data collected at $\sqrt{s}=$ 4.599 $\gev$ with the BESIII detector, we present the first observation of the SCS decays $\lambdacp\to\ppipi$ and improved (or comparable) measurements of the $\lambdacp\to p\phi$ and $\lambdacp\to\pkk_{\text{non-}\phi}$ BFs comparing to PDG values~\cite{2014PDG}.
The relative BFs with respect to the CF decay $\lambdacp\to\pkpi$ are measured.
Taking $\BR(\lambdacp\to\pkpi)=(5.84 \pm 0.27 \pm 0.23)\%$ from Ref.~\cite{lambdcpBr}, we also obtain absolute BFs for the SCS decays.
All the results are summarized in Table~\ref{tab:br}.
The results provide important data to understand the dynamics of
	$\lambdacp$ decays.
	They especially help to
    distinguish predictions from different theoretical models and understand contributions
	from factorizable effects~\cite{2015Yhaiyang}.

The BESIII Collaboration thanks the staff of BEPCII, the IHEP computing center and the supercomputing center
of USTC for their strong support.
This work is supported in part by National Key Basic Research Program of China under
Contract No. 2015CB856700; National Natural Science Foundation of China (NSFC) under Contracts
No. 11125525, No. 11235011, No. 11322544, No. 11335008, No. 11425524, No. 11322544, No. 11375170, No. 11275189,  No. 11475169, No. 11475164;
the Chinese Academy of Sciences (CAS) Large-Scale Scientific Facility Program; the CAS
Center for Excellence in Particle Physics (CCEPP); the Collaborative Innovation Center for Particles
and Interactions (CICPI); Joint Large-Scale Scientific Facility Funds of the NSFC and CAS under
Contracts No. 11179007, No. U1232201, No. U1332201;
CAS under Contracts No. KJCX2-YW-N29 and No. KJCX2-YW-N45;
100 Talents Program of CAS; INPAC and Shanghai Key Laboratory for Particle Physics and Cosmology;
German Research Foundation DFG under Collaborative Research Center CContract No. RC-1044, FOR 2359; Istituto
Nazionale di Fisica Nucleare, Italy; Ministry of Development of Turkey under Contract
No. DPT2006K-120470; Russian Foundation for Basic Research under Contract No. 14-07-91152;
U. S. Department of Energy under Contracts No. DE-FG02-04ER41291, No. DE-FG02-05ER41374,
No. DE-FG02-94ER40823, No. DESC0010118; U.S. National Science Foundation; University of Groningen (RuG) and
the Helmholtz Centre for Heavy Ion Research GmbH (GSI), Darmstadt; and WCU Program of National Research
Foundation of Korea under Contract No. R32-2008-000-10155-0.

\end{document}